\documentclass[
reprint, amsmath,amssymb,aps,prl,superscriptaddress
]{revtex4-1}

\usepackage{graphicx}
\usepackage{amsmath} 
\usepackage[colorlinks,urlcolor=blue,linkcolor=blue,citecolor=blue]{hyperref}
\usepackage{amsfonts}
\usepackage{bbm}
\usepackage{braket}
\usepackage{graphicx}
\usepackage{float}
\usepackage{url}
\usepackage{qcircuit}
\usepackage{amssymb}
\usepackage{qcircuit}
\usepackage{siunitx}
\usepackage{braket}
\usepackage{multirow}
\usepackage{color}
\usepackage{gensymb}

\newcommand{\dqc}{Duke Quantum Center, Duke University, Durham, NC 27701, USA}

\begin{document}

\preprint{APS/123-QED}

\title{Realization of Scalable Cirac-Zoller Multi-Qubit Gates}

\author{Chao Fang}
\email{chao.fang@duke.edu}
\affiliation{\dqc}
\affiliation{Department of Electrical and Computer Engineering, Duke University, Durham, NC 27708, USA}
\author{Ye Wang}
\altaffiliation{Current address: School of Physical Sciences, University of Science and Technology of China, Hefei 230026, China}
\affiliation{\dqc}
\affiliation{Department of Electrical and Computer Engineering, Duke University, Durham, NC 27708, USA}
\author{Ke Sun}
\affiliation{\dqc}
\affiliation{Department of Physics, Duke University, Durham, NC 27708, USA}
\author{Jungsang Kim}
\email{jungsang.kim@duke.edu}
\affiliation{\dqc}
\affiliation{Department of Electrical and Computer Engineering, Duke University, Durham, NC 27708, USA}
\affiliation{Department of Physics, Duke University, Durham, NC 27708, USA}
\affiliation{IonQ, Inc., College Park, MD 20740, USA}

\date{\today}


\maketitle

\noindent\textbf{The universality theorem in quantum computing states that any quantum computational task can be decomposed into a finite set of logic gates operating on one and two qubits \cite{Deutsch1995}. However, the process of such decomposition is generally inefficient, often leading to exponentially many gates to realize an arbitrary computational task. Practical processor designs benefit greatly from availability of multi-qubit gates that operate on more than two qubits to implement the desired circuit. In 1995, Cirac and Zoller proposed a method to realize native multi-qubit controlled-$Z$ gates in trapped ion systems, which has a stringent requirement on ground-state cooling of the motional modes utilized by the gate \cite{Cirac1995}. An alternative approach, the M\o lmer-S\o rensen (MS) gate \cite{sorensen1999quantum}, is robust against residual motional excitation and has been a foundation for many high-fidelity gate demonstrations \cite{ballance2016high, gaebler2016high, clark2021high, srinivas2021high}. This gate does not scale well beyond two qubits, incurring additional overhead when used to construct many target algorithms \cite{Vedral1996, Draper2006, Grover1996, Wang2001, Figgatt2017, Nam2020, Cory1998, Reed2012, Paetznick2013}. Here, we take advantage of novel performance benefits of long ion chains to realize fully programmable and scalable high-fidelity Cirac-Zoller gates.}

A prototypical example of useful multi-qubit gates is the $N$-qubit Toffoli gate, which performs a conditional bit-flip (Pauli-$X$) on the target qubit if and only if all other $N-1$ control qubits are in the state $\ket{1}$. Such a gate can be realized by a more symmetric $N$-qubit conditional phase-flip gate where the state picks up a minus sign when all $N$ qubits are in $\ket{1}$, combined with Hadamard gates applied to the target qubit \cite{NielsenChuang}. Realization of the Toffoli gate was demonstrated in physical platforms including nuclear magnetic resonance \cite{Cory1998}, trapped ion \cite{Monz2009, Figgatt2017}, superconducting circuit \cite{Fedorov2012, Reed2012, Hill2021, Kim2022}, linear optics \cite{Micuda2013, Micuda2015} and neutral atom \cite{Levine2019} systems with up to $N=4$ qubits \cite{Figgatt2017}. When implemented using the traditional decomposition into controlled-NOT (CNOT) gates and single-qubit rotations, the $N$-Toffoli gate carries a quadratic overhead in general \cite{Shende2019}, or a linear overhead when ancilla qubits are introduced \cite{Maslov2003, He2017, Yu2013}. Efficient implementation of $N$-Toffoli gates is essential in important applications such as elementary arithmetic \cite{Vedral1996, Draper2006}, Grover search \cite{Grover1996, Wang2001, Figgatt2017}, quantum Fourier transform \cite{Nam2020} and quantum error correction algorithms \cite{Cory1998, Reed2012, Paetznick2013}. 

Instead of decomposing into two-qubit gates, methods utilizing native interactions between multiple qubits have been proposed to directly implement the $N$-Toffoli gate \cite{Wang2001, Cirac1995, Goto2004, Rasmussen2020, Levine2019, Isenhower2011, Katz2022}. Most notable is the Cirac-Zoller gate scheme, the first experimental protocol for entangling qubits using trapped ions \cite{Cirac1995, Monz2009}. In this scheme, the spin states of individual ions are coupled to a collective motional degree of freedom, which serves as an information ``bus'' to mediate the interactions between multiple spins. The conditional operation is realized by the blockade of the motional ground state of the bus mode, which allows a phonon to be added but not extracted. As a result, the fidelity of the Cirac-Zoller gate is directly limited by how well the collective motion can be initially cooled to the ground state $\ket{n=0}$, and the subsequent heating of the ions during the gate. In this letter, we report the demonstration of the $N$-Toffoli gate with up to $N=5$ in a five-ion chain. We overcome traditional experimental challenges of the Cirac-Zoller gate with three innovations. First, we note that for normal modes of motion in a linear ion chain, the technical noise that heats up the center-of-mass (COM) mode barely affects the lower-frequency modes. By utilizing non-COM modes, we can cool and maintain the ion chain well in its ground state throughout the gate operation. Second, the gate scheme requires an auxiliary state to temporarily store the quantum information for participating qubits. Our system features extended coherence times for the Zeeman states in the ground state manifold of the ions, sufficient to serve as the auxiliary state, achieved by magnetic field stabilization. Lastly, our system enables individual addressing of qubits with low crosstalk error, which is a major limitation in previous implementations \cite{Monz2009}. These capabilities allow us to realize the Cirac-Zoller gate with performance competitive with other gate schemes.

\begin{figure*}[htbp]
    \centering
    \includegraphics[width=\linewidth]{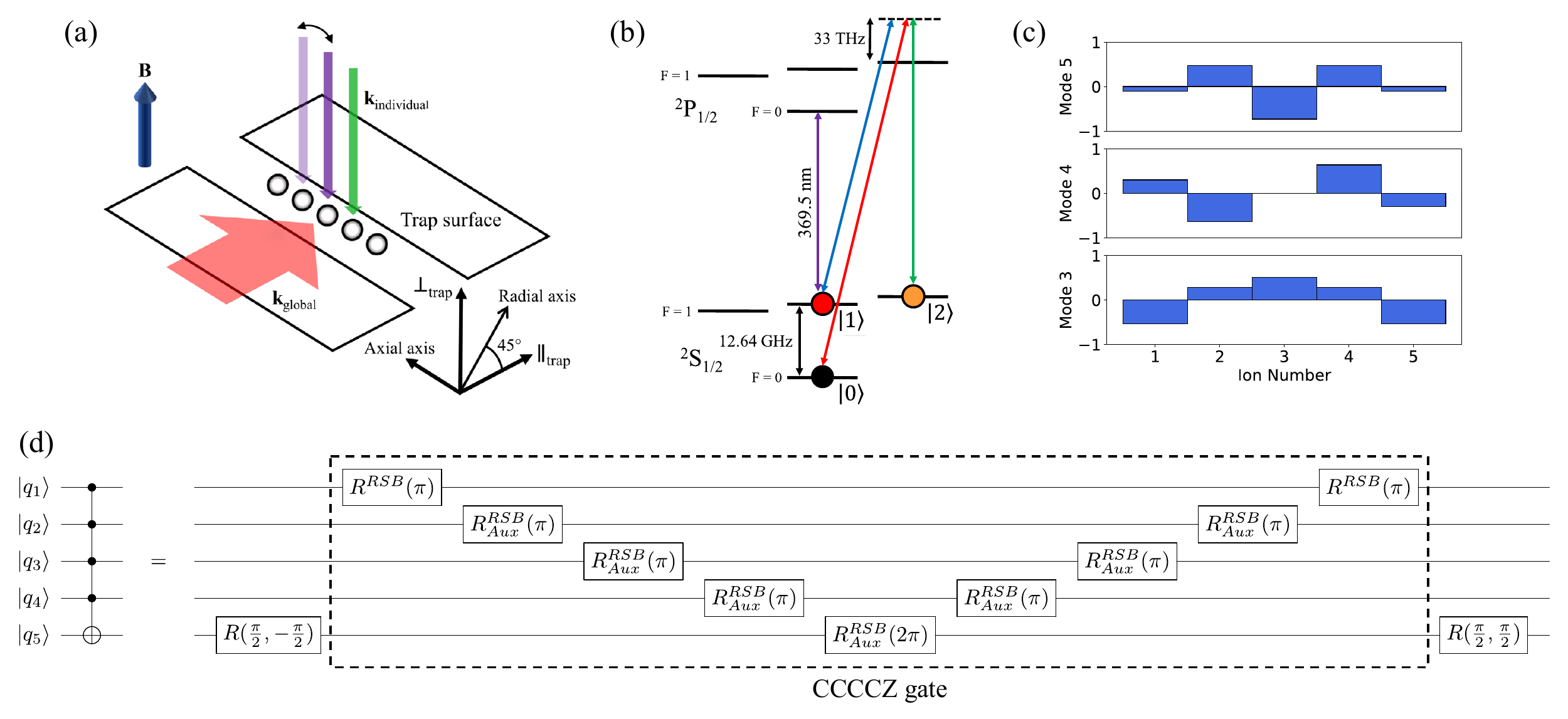}
    \caption{
    {{(a) Schematic representation of the experimental setup. A chain of five ions is confined in a surface trap and addressed by three Raman beams consisting of two individual addressing beams and an orthogonal global beam whose wave vectors couple optimally to the radial modes of motion along the 45\degree axis. The individual addressing beams can be steered by microelectromechanical system mirrors. (b) Energy diagram of the relevant atomic levels of $^{171}$Yb$^{+}$ ion with spin and motional state manipulations driven by stimulated Raman transitions. The hyperfine clock levels encode the qubit states $\ket{0}$ and $\ket{1}$ while one of the Zeeman levels serves as the auxiliary state $\ket{2}$ required by the Cirac-Zoller protocol. (c) Coupling strength of each ion to the fifth (zig-zag), the fourth and the third radial motional modes. We choose these three modes in order as the bus modes to implement the 3-, 4- and 5-Toffoli gates, respectively. For the 3- and 4-Toffoli gates, we use the ions that have stronger coupling to the respective bus mode to reduce pulse times, i.e. the three center ions for the 3-Toffoli and the four outer ions for the 4-Toffoli. (d) Circuit for implementing the 5-Toffoli gate using the Cirac-Zoller protocol. The CCCCZ pulse sequence is sandwiched between two $\pi/2$ rotations on the target qubit to result in a 5-Toffoli gate. Here the numbering of qubits has no correspondence with the ion number in Fig. 1(c).}}}
    \label{config}
\end{figure*}

The quantum device in this work is based on a chain of five $^{171}$Yb$^+$ ions confined in a microfabricated linear Paul trap \cite{Revelle2020}, as illustrated in Fig. 1(a). The qubit spin states are represented by the hyperfine ``clock'' levels in the $^2$S$_{1/2}$ ground state of each ion $\ket{0} \equiv \ket{F=0;m_F = 0}$ and $\ket{1} \equiv \ket{F=1;m_F = 0}$, as shown in Fig. 1(b). We drive stimulated Raman transitions using a set of three pulsed laser beams to manipulate the qubit spin and motional states: two individual addressing beams and an orthogonal global beam whose wave vectors couple optimally to the radial modes of motion along the 45\degree axis shown in Fig. 1(a). The individual addressing beams can be steered by microelectromechanical system (MEMS) mirrors \cite{crain2014} to address each ion with a gate crosstalk of $\sim 2\%$ on the nearest neighbor ions \cite{Fang2022}. Detailed description of the experimental setup is provided in Ref. \cite{wang2020}. 

In the Cirac-Zoller protocol, we drive red sideband (RSB) transition to couple the spin states $\ket{0}$ and $\ket{1}$ to the motional Fock state $\ket{n}$ of a bus mode. In the Lamb-Dicke regime where the amplitude of motion is small compared to the laser wavelength, the interaction is described by the Hamiltonian
\begin{align}\label{eq:H}
    \hat{H}_{RSB}(t) &= i\Omega_r\left[\hat{\sigma}_+\hat{a} e^{-i(\delta t+\phi_r)}+\hat{\sigma}_-\hat{a}^\dagger e^{i(\delta t+\phi_r)}\right],
\end{align}
where $\hat{\sigma}_{+(-)}$ is the spin raising (lowering) operator, $\hat{a}^\dag(\hat{a})$ is the phonon creation (annihilation) operator, $\delta$ is the laser detuning, and $\Omega_{r}$ and $\phi_{r}$ are the Rabi frequency and phase of RSB transition. If we choose $\delta=0$ and $\phi_{r}=0$, a resonant RSB $\pi$ pulse $R^{RSB}(\pi)$ can induce a spin bit-flip together with a $\pi/2$ phase shift depending on the initial spin and motional states, written as
\begin{align}
    \begin{aligned}
    \ket{1}\ket{n=0}&\xrightarrow{RSB}-i\ket{0}\ket{n=1}\xrightarrow{RSB}-\ket{1}\ket{n=0},\\
    \ket{0}\ket{n=0}&\xrightarrow{RSB}\ket{0}\ket{n=0}.
    \end{aligned}\label{eq:rsb}
\end{align}
When initialized to $\ket{0}\ket{n=0}$, no transition can be driven by $R^{RSB}(\pi)$ due to the ground state blockade. The Cirac-Zoller protocol requires a second RSB transition between $\ket{0}$ and an auxiliary state $\ket{2}$. This can be circumvented for the cases of CNOT gate \cite{Schmidt-Kaler2003} and 3-Toffoli gate \cite{Monz2009} by designing clever composite pulse sequences. However, these approaches do not scale to $N>3$. In our experiment, we use one of the Zeeman levels in the $^2$S$_{1/2}$ ground state $\ket{F=1;m_F =\pm 1}$ as the auxiliary state $\ket{2}$, as shown in Fig. 1(b) (see Methods for details). Similar to Eq.~(\ref{eq:rsb}), the action of the $\pi$ pulse $R_{Aux}^{RSB}(\pi)$ is conditional on the initial spin and motional states, allowing only the transition
\begin{align}
    \ket{0}\ket{n=1}\xrightarrow[Aux]{RSB}-i\ket{2}\ket{n=0}\xrightarrow[Aux]{RSB}-\ket{0}\ket{n=1}.\label{eq:rsb2}
\end{align}
Now we look at the action of the pulse sequence
\begin{align}\label{eq:U}
    U = &R_1^{RSB}(\pi) \prod_{j=N-1}^2 R_{j,Aux}^{RSB}(\pi) R_{N,Aux}^{RSB}(2\pi) \prod_{j=2}^{N-1} R_{j,Aux}^{RSB}(\pi)\nonumber \\ 
    \times & R_1^{RSB}(\pi),
\end{align}
where $R_j^{RSB}(\pi)$ [$R_{j,Aux}^{RSB}(\pi)$] is the RSB $\pi$ pulse for the clock (Zeeman) transition on ion $j$. Following Eqs.~(\ref{eq:rsb}) and~(\ref{eq:rsb2}), it is straightforward to see that only one computational basis state undergoes a $\pi$ phase shift $\ket{11...1}\ket{n=0}\rightarrow-\ket{11...1}\ket{n=0}$ with all the other basis states unchanged, which is exactly the $N$-qubit controlled-$Z$ unitary operation on the spin state. The $N$-Toffoli gate is then achieved by applying a Hadamard gate on the target qubit before and after the sequence of Eq.~(\ref{eq:U}). Figure 1(d) shows the circuit for the 5-Toffoli gate with qubit 5 chosen as the target, which can be any of the five qubits in the chain.

\begin{figure*}[htbp]
    \centering
    \includegraphics[width=0.9 \linewidth]{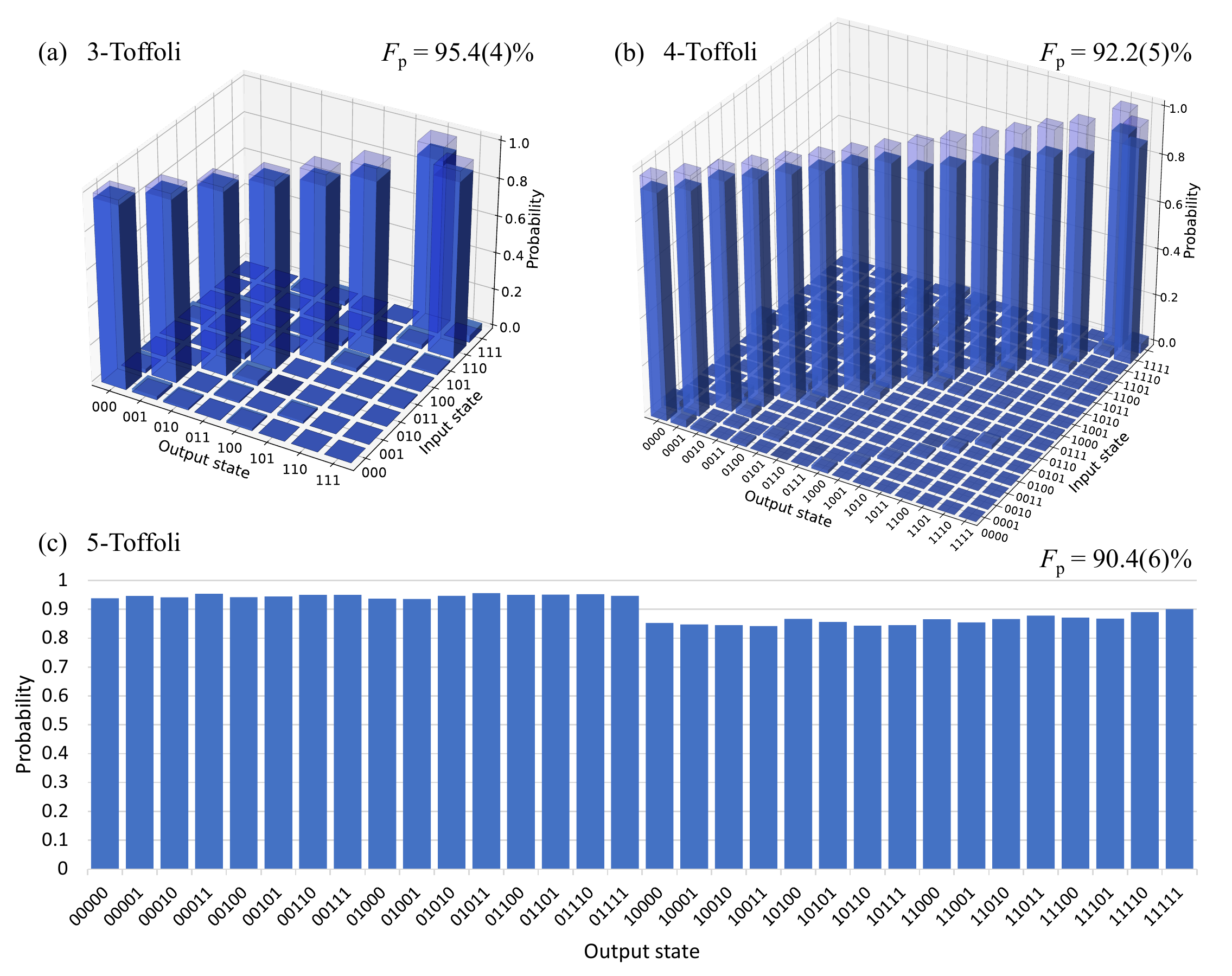}
    \caption{
    {{Experimentally obtained truth tables for the (a) 3-Toffoli gate, (b) 4-Toffoli gate, and (c) 5-Toffoli gate. The wire frames show the ideal outcomes while the solid bars represent the measured population distributions for all the computational basis states. The average populations of correct output states $F_p$ are $95.4(4)\%$, $92.2(5)\%$ and $90.4(6)\%$ for the 3-, 4- and 5-Toffoli gates, respectively, corrected for state preparation and measurement (SPAM) errors ($\sim 0.5\%$ per ion).}}}
    \label{config}
\end{figure*}

Despite having equal coupling to all ions in a chain, the COM motional mode typically suffers from a large heating rate due to its susceptibility to uniform electric field noise \cite{Wineland1998}. In our system, we measure the heating rates to be 614(18) quanta/s for the radial COM mode and $<10$ quanta/s for the other radial modes in a five-ion chain. Therefore we choose the fifth (numbered from highest to lowest mode frequencies, also known as the ``zig-zag'' mode), the fourth and the third radial modes as the bus modes to implement the 3-, 4- and 5-Toffoli gates, respectively. For the 3- and 4-Toffoli gates, we use the ions that have stronger coupling to the respective bus mode to reduce pulse times, as shown in Fig. 1(c). The zig-zag and the fourth modes are preferable because of their larger frequency separation from the fast heating COM mode, but one can use the third mode to implement all three gates for arbitrary selection of ions with a minimal performance penalty. 

At the beginning of each experiment, the ion chain is laser cooled using Doppler and electromagetically-induced-transparency cooling \cite{Feng2020, Qiao2021} to $\bar{n}\approx 0.5$ for all five modes, then the bus mode is further cooled to the ground state {($\bar{n}\approx0.02$)} via resolved-sideband cooling. We prepare the system in each of the computational basis states for $N$ qubits. After applying the Cirac-Zoller gate pulse sequence, we characterize the gate performance by measuring the output state distribution for each of the input eigenstates, to map out the truth table (see Methods for details). Figure 2 shows the measured truth tables as bar graphs and the average population fidelities $F_p$ for $N=3$, 4 and 5. For the 5-Toffoli gate, only the correct output state populations are shown in the bar graph (Fig. 2(c)). As the pulse sequence only increases by two $R_{Aux}^{RSB}(\pi)$ rotations from $N$ to $N+1$ qubits, the fidelity drop with larger $N$ is much slower compared with the Toffoli gate implementation based on two-qubit gates \cite{Figgatt2017}.

The truth table in the computational basis represents the magnitude of the matrix elements of the Toffoli unitary operator, but carries no information on the relative phases between the different entries, which is necessary to verify the quantum operation of the gate. Quantum process tomography \cite{Riebe2006} provides a complete characterization of quantum operations and has been used to reconstruct the full process matrix of 3-Toffoli gates \cite{Monz2009, Fedorov2012, Reed2012, Hill2021, Kim2022}, but the number of measurement settings required grows exponentially large ($16^N-4^N$), and quickly becomes experimentally impractical. A truth table measurement with input states rotated to the $X$ computational basis can recover some phase information, and complement the $Z$ basis truth table measurement to provide a lower bound estimation of the process fidelity \cite{Hofmann2005}. A restricted version of this procedure dubbed ``limited tomography'' has been used to verify the phases of the 3-Toffoli gate unitary \cite{Figgatt2017, Levine2019}. 

We use a method based on a generalization of the fidelity bound in Ref. \cite{Hofmann2005} to estimate the upper and the lower bounds for the process fidelity of our $N$-Toffoli gates \cite{Micuda2013}. We prepare $N$ sets of input states in the computational basis but with the spin of ion $k$ ($k=1,2,...N$) rotated to $\ket{\pm} \equiv \left(\ket{0}\pm\ket{1}\right)/\sqrt{2}$ and apply the $N$-qubit controlled-$Z$ gate which produces output states within the same basis of each set of input states. The truth table fidelity $F_k$ for each set can be obtained via simple projection measurements, and the process fidelity $F_\chi$ can be estimated as 
\begin{align}
    \sum_{k=1}^{N}F_k-N+1 \leq F_\chi \leq \min(F_k).\label{eq:F}
\end{align}
Similar to the limited tomography procedure, the fidelity $F_k$ verifies that the phase is implemented correctly for ion $k$. The measured truth tables and the $F_k$ values for $N=3$ (the controlled-controlled-$Z$ or CCZ gate) are shown in Fig. 3. With these we estimate the process fidelity bounds $78.8\% \leq F_\chi \leq 92.4\%$. The truth table measurements with partially conjugate basis states are repeated for $N=4$ and 5, with the results shown in Extended Data Figs. 2 and 3. We note that even though the decrease in fidelity $F_k$ for each set of input states is moderate as $N$ increases, the lower bound estimation of $F_\chi$ quickly drops below $50\%$ due to the number of $F_k$ measurements required.

\begin{figure*}[htbp]
    \centering
    \includegraphics[width=\linewidth]{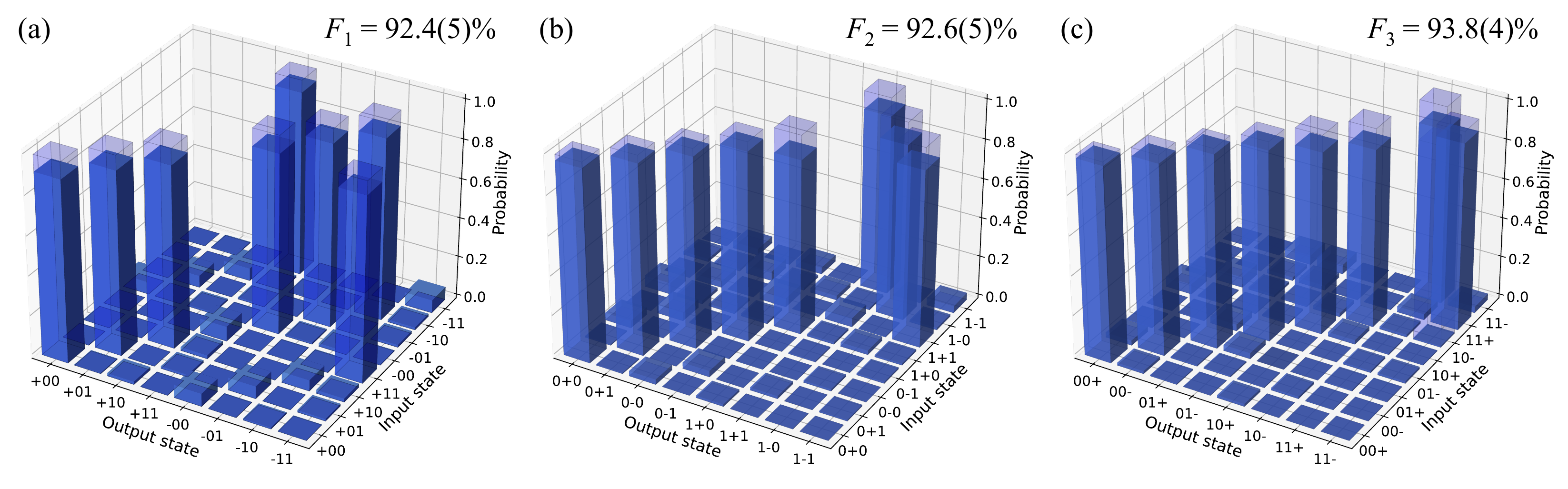}
    \caption{Experimentally obtained truth tables for the CCZ gate in three partially conjugate bases with ion 1 (a), ion 2 (b) and ion 3 (c) initialized to $\ket{\pm}$. The measured truth table fidelities are $F_1=92.4(5)\%$, $F_2=92.6(5)\%$ and $F_3=93.8(4)\%$, corrected for SPAM errors, which can be used to verify that the implemented CCZ gate has no spurious phases on the expected output states, and estimate the upper and the lower bounds for the process fidelity.}
    \label{diag}
\end{figure*}

The effect of various error sources on the measured Toffoli gate truth table fidelity in the computational basis are analyzed using numerical simulation. The contribution of each individual error source is calculated by solving the master equation in the Lindblad form, following the method described in the supplementary material of Ref. \cite{wang2020}. The simulated error budget is shown in Table~\ref{tab:error}. Only the resonant bus mode is considered for the 4-Toffoli gate case due to the high computational cost demanded by the size of the full density matrix when the off-resonant modes are included in the simulation. For the 3-Toffoli gate, motional heating, imperfect ground state cooling and motional dephasing of the bus mode each causes $\sim 1\%$ error. The motional coherence time is measured to be 8(1) ms in our system. We consider two off-resonant modes (the spectrally closest fourth mode and the fast heating COM mode) in the full density matrix to reduce computational cost, and assume that the errors from heating and imperfect ground state on the off-resonant modes are both additive to the effect of off-resonant mode coupling alone. Despite a significantly higher heating rate, the COM mode causes a negligible error thanks to its large frequency separation from the bus mode. Surprisingly, the initial temperature of the off-resonant fourth mode ($\bar{n}\approx 0.5$) has the largest error contribution, which may explain the steeper fidelity drop for the 4-Toffoli gate since its bus mode has two nearby off-resonant modes, same as the 5-Toffoli gate case.

\begin{table}[!htb]
\setlength{\tabcolsep}{6.5pt}
\renewcommand{\arraystretch}{1.5}
\centering
\begin{tabular}{c|c|c}
\hline
\multirow{2}{*}{Error source} & 3-Toffoli & 4-Toffoli \\
&error ($\%$) & error ($\%$) \\
\hline
Motional heating (bus mode) & $0.96 \pm 0.19$ & $1.56 \pm 0.31$ \\
Initial $\bar{n}$ (bus mode) & $0.89 \pm 0.44$ & $0.73 \pm 0.36$ \\
Motional dephasing (bus mode) & $0.94 \pm 0.13$ & $1.24 \pm 0.18$ \\
Zeeman dephasing & $0.07 \pm 0.05$ & $0.07 \pm 0.05$ \\
Addressing crosstalk & $0.41 \pm 0.16$ & $0.46 \pm 0.16$ \\
Off-resonant mode coupling & $0.57$ \\
Heating (off-resonant modes) & $0.06 \pm 0.01$ \\
Initial $\bar{n}$ (off-resonant modes) & $1.4 \pm 0.31$ \\
Total & $5.31 \pm 0.61$ \\
\hline
Experiment total & $4.6 \pm 0.4$ & $7.8 \pm 0.5$\\
\hline
\end{tabular}
\caption{Simulated error budget and experimental measurement of the truth table fidelity in the computational basis for 3- and 4-Toffoli gates. Only the bus mode is considered for the 4-Toffoli gate case due to the computational cost of the full density matrix size including the off-resonant modes.
}\label{tab:error}
\end{table}

In conclusion, the Cirac-Zoller gate protocol uses a simple pulse sequence to scalably realize the $N$-Toffoli gate, compared with circuit decomposition approach into two-qubit gates. We report the implementation of the $N$-Toffoli gate with up to $N=5$ qubits for the first time in a trapped-ion system featuring long-coherence Zeeman levels that are suitable to be used as the auxiliary state necessary for the protocol, thereby demonstrating that the Cirac-Zoller gate is an attractive solution for running many complex circuits in trapped-ion quantum computers. Our gate time is limited by the available power and polarization configuration of the Raman beams, allowing for a maximum sideband Rabi frequency of $\sim 5$ kHz. A faster gate can reduce error caused by heating and decoherence, but at the cost of increasing the unwanted coupling to off-resonant motional modes. Looking forward, pulse modulation to decouple all the off-resonant modes at the end of each RSB $\pi$ pulse similar to the techniques used in MS gates \cite{Zhu2006, Roos_2008, Leung2018} can be employed to achieve high gate fidelities. The Cirac-Zoller gate can also benefit from newer cryogenic systems that feature extremely low heating rates \cite{Spivey2022}. With an implementation that can be readily adopted and further scaled, the $N$-Toffoli gate is certain to find wider uses in quantum information applications.
 
\section{References}
\bibliographystyle{apsrev4-1} 
\bibliography{./Reference}

\section{Methods}

\subsection{Using a Zeeman level as the auxiliary state $\ket{2}$} 
The Zeeman levels in the $^2$S$_{1/2}$ ground state $\ket{F=1;m_F =\pm 1}$ of $^{171}$Yb$^+$ ions are first-order-sensitive to magnetic field fluctuations and therefore susceptible to decoherence. Without magnetic field stabilization, we typically measure a Zeeman state coherence time $<1$ ms in our system. To reduce gate error caused by Zeeman state decoherence, we use permanent magnets to generate the quantization field and shield the vacuum chamber with a dual-layer mu-metal enclosure which provides over 20 dB attenuation of magnetic field noise. The Zeeman state coherence time is increased to $>100$ ms, characterized by Ramsey interferometry measurement (Extended Data Fig. 1), and does not limit our gate performance. Further improvement is possible with an AC line trigger \cite{Ruster2016}.

The Zeeman level transition between $\ket{0}$ and $\ket{2}$ requires $\pi$-polarization in the Raman beams, which is usually eliminated to maximize the Rabi frequency of clock transition for Raman gates. Since our system does not support fast polarization control of the Raman beams between gate pulses, and the two individual addressing beams share a polarization setting due to the optical setup, we cannot drive the $\ket{0}$ to $\ket{1}$ transition and the $\ket{0}$ to $\ket{2}$ transition using separate polarizations as proposed in the original Cirac-Zoller protocol. Instead, a polarization configuration is chosen to compromise between the Rabi frequencies of clock and Zeeman transitions, with a maximum sideband Rabi frequency of $\sim 5$ kHz for both transitions.

\subsection{Measurement protocol for accurate $\ket{1}$ populations} 
 We note that depending on the input state, some spins can temporarily populate $\ket{2}$ during the gate, which is outside the computational subspace, and residual population in $\ket{2}$ at the end of the gate is possible due to imperfect RSB $\pi$-rotations. As the projection measurement using spin-dependent fluorescence in our system cannot distinguish between $\ket{1}$ and $\ket{2}$, we apply an additional carrier $\pi$ rotation on each ion at the end, which flips $\ket{0}$ and $\ket{1}$ populations without affecting $\ket{2}$, to obtain the accurate output state populations. All single-qubit gates are implemented using SK1 composite pulse sequence \cite{brown2004sk1}.

\section{Acknowledgements}
We thank Bichen Zhang for his contribution to the experiments and Zhubing Jia for helpful discussions. This work is supported by the Office of the Director of National Intelligence - Intelligence Advanced Research Projects Activity through ARO contract W911NF-16-1-0082 and the DOE BES award DE-SC0019449.

\onecolumngrid
\newpage
\section{Extended Data}
\setcounter{figure}{0}
\renewcommand{\figurename}{Extended Data FIG.}

\begin{figure}[htbp]
\center{\includegraphics[width=0.56 \textwidth]{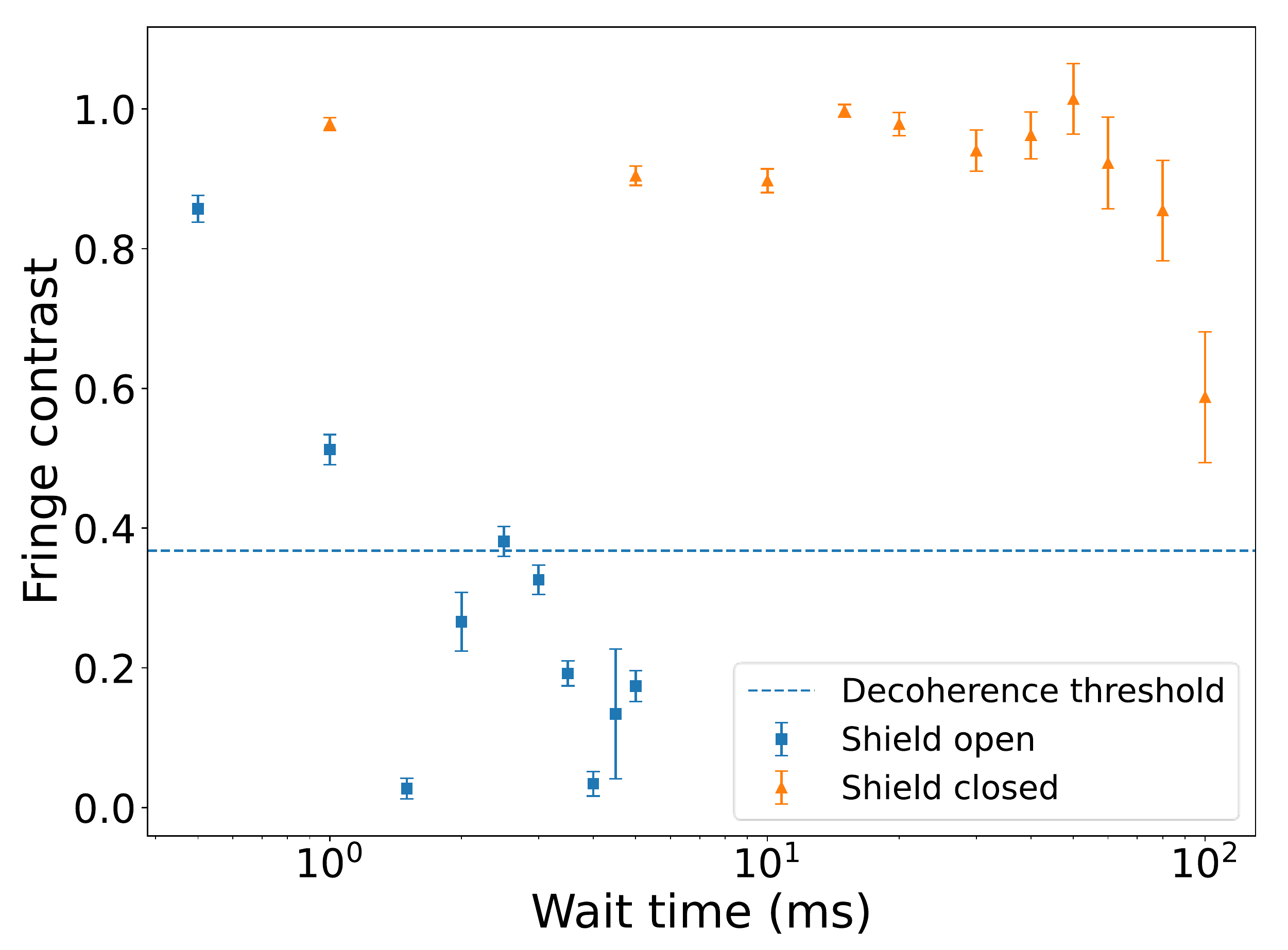}}
\caption{Ramsey interferometry measurement characterizing the Zeeman state coherence time. The results with the mu-metal shield in an open configuration and full closed are compared. The Ramsey fringe contrast as a function of the wait time indicates a Zeeman state coherence time $>100$ ms with the shield closed, which does not limit the gate fidelity.}
\end{figure}

\newpage

\begin{figure}[htbp]
\center{\includegraphics[width=1 \textwidth]{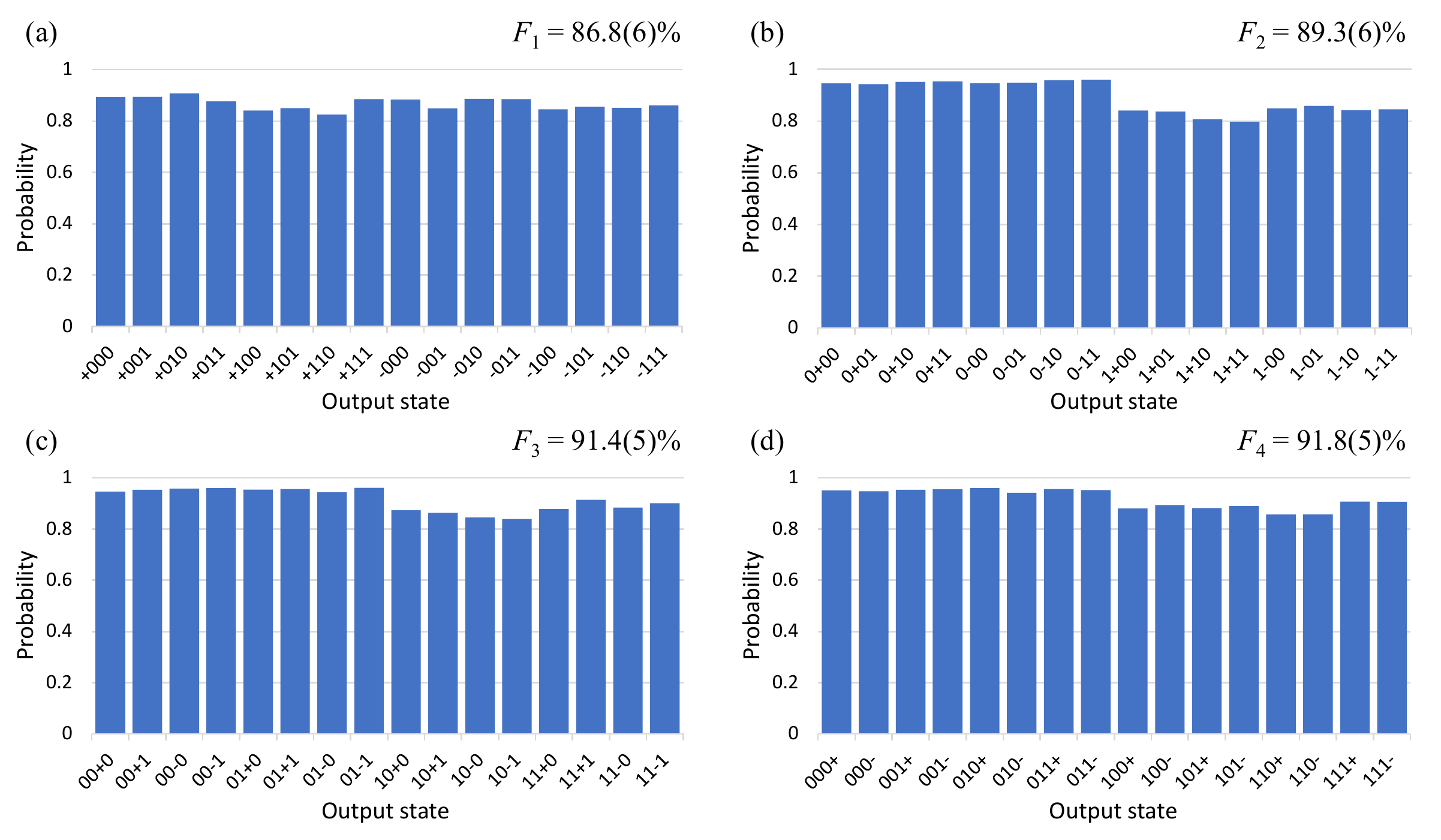}}
\caption{Correct output state populations for the CCCZ gate in four partially conjugate bases with ion 1 (a), ion 2 (b), ion 3 (c) and ion 4 (d) initialized to $\ket{\pm}$. The measured truth table fidelities are $F_1=86.8(6)\%$, $F_2=89.3(6)\%$, $F_3=91.4(5)\%$ and $F_4=91.8(5)\%$, corrected for SPAM errors, which can be used to verify that the implemented CCCZ gate has no spurious phases on the expected output states, and estimate the upper and the lower bounds for the process fidelity to be $59.3\% \leq F_\chi \leq 86.8\%$.}
\end{figure}

\newpage

\begin{figure}[htbp]
\center{\includegraphics[width=1 \textwidth]{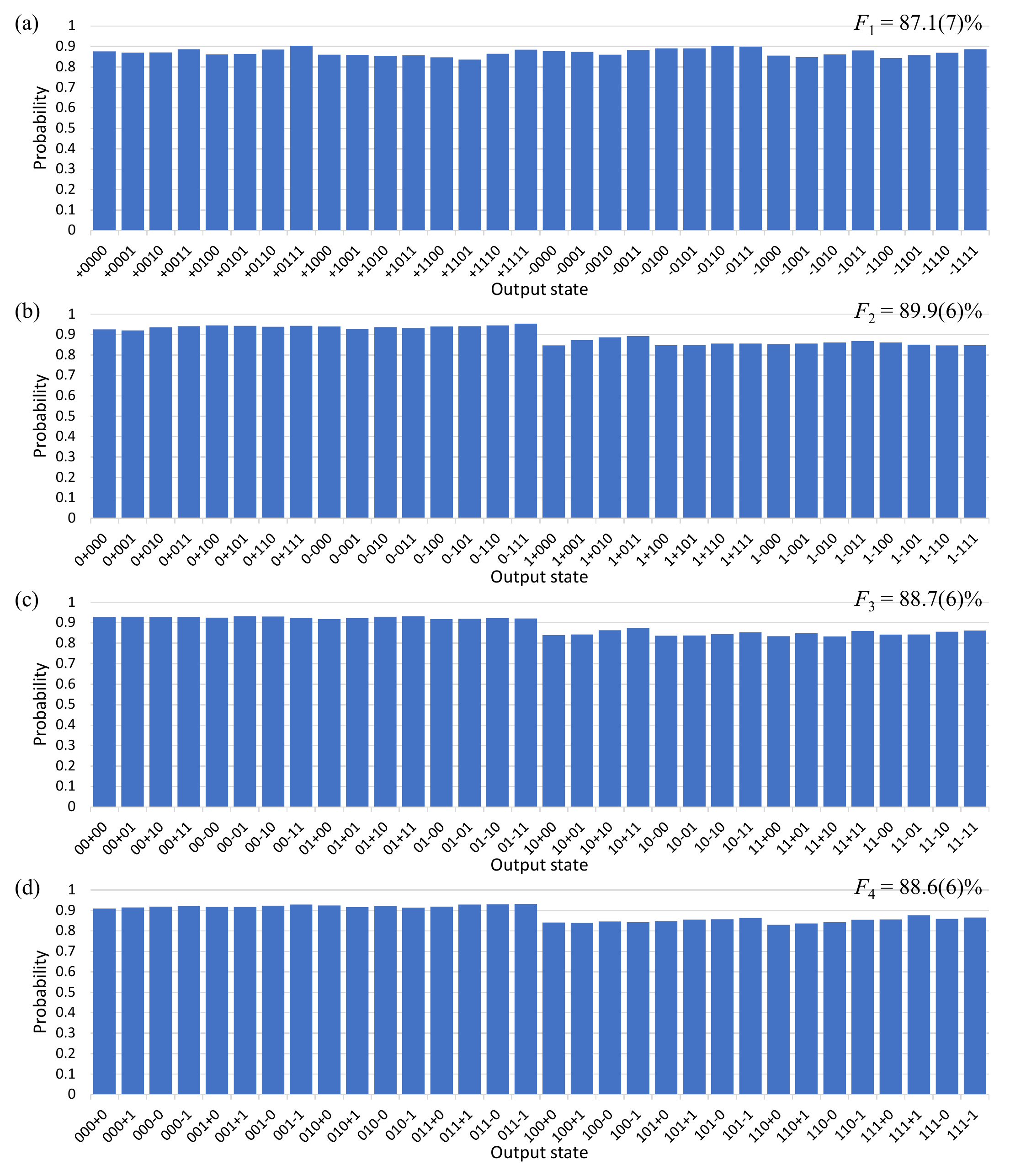}}
\end{figure}

\newpage

\begin{figure}[htbp]
\center{\includegraphics[width=1 \textwidth]{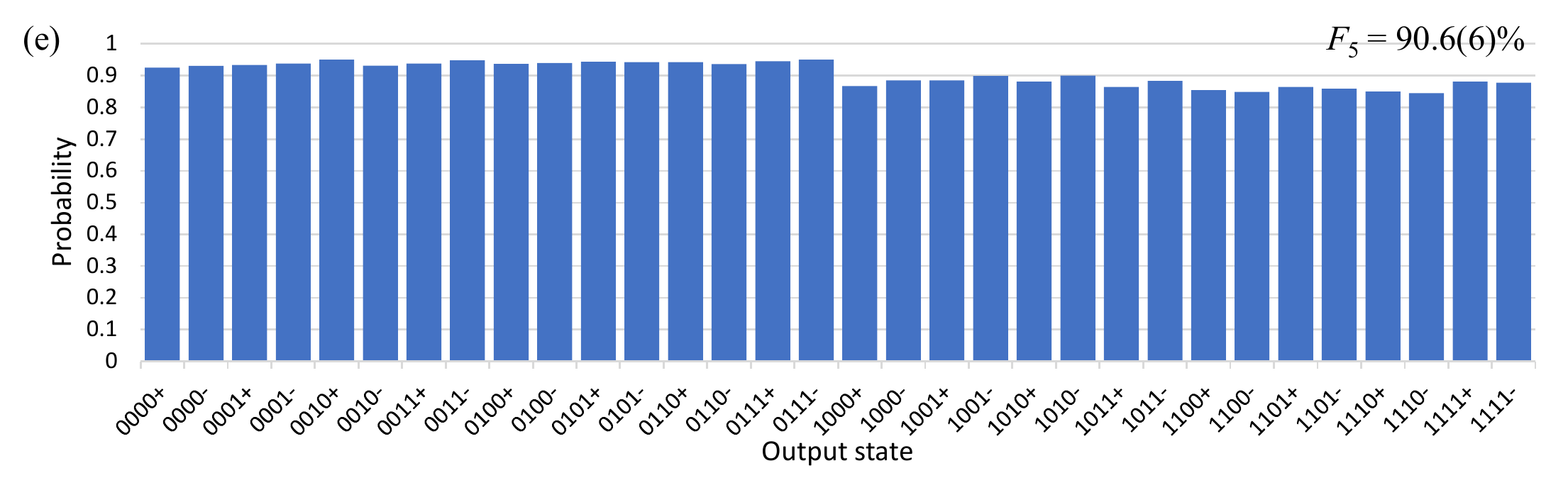}}
\caption{Correct output state populations for the CCCCZ gate in five partially conjugate bases with ion 1 (a), ion 2 (b), ion 3 (c), ion 4 (d) and ion 5 (e) initialized to $\ket{\pm}$. The measured truth table fidelities are $F_1=87.1(7)\%$, $F_2=89.9(6)\%$, $F_3=88.7(6)\%$, $F_4=88.6(6)\%$ and $F_5=90.6(6)\%$, corrected for SPAM errors, which can be used to verify that the implemented CCCCZ gate has no spurious phases on the expected output states, and estimate the upper and the lower bounds for the process fidelity to be $44.9\% \leq F_\chi \leq 87.1\%$.}
\end{figure}

\end{document}